\documentclass[12pt]{article}

\usepackage{amstext,amsmath,amssymb,amsfonts,bbm}
\usepackage[latin1]{inputenc}
\usepackage{epsfig}
\usepackage{hyperref}
\usepackage{amsthm}
\usepackage{subfigure}
\usepackage{color}
\usepackage{multirow}
\newcommand{\bea}{\begin{eqnarray}}	
\newcommand{\eea}{\end{eqnarray}}
\newcommand{\be}{\begin{equation}}	
\newcommand{\ee}{\end{equation}}
\newcommand{\cG}{{\cal G}}
\newcommand{\cV}{{\cal V}}
\newcommand{\cC}{{\cal C}}
\newcommand{\cF}{{\cal F}}
\newcommand{\cB}{{\cal B}}
\newcommand{\cL}{{\cal L}}
\newcommand{\cH}{{\cal H}}
\newcommand{\mB}{\mathfrak{B}}
\newcommand{\ZZ}{\mathbb{Z}}

\newtheorem{lemma}{Lemma}

\newtheorem{definition}{Definition}

\begin{document}

\title{Topological Graph Polynomials in Colored Group Field Theory}

\author{Razvan Gurau\footnote{Perimeter Institute for Theoretical Physics Waterloo, ON, N2L 2Y5, Canada. } }

\date{}

\maketitle

\begin{abstract}
\noindent In this paper we analyze the open Feynman graphs of the Colored Group Field Theory introduced in \cite{Color}. We define the boundary graph $\cG_{\partial}$ of an open graph $\cG$ and prove it is a cellular complex. Using this structure we generalize the topological (Bollob\'as-Riordan) Tutte polynomials associated to (ribbon) graphs to topological polynomials adapted to Colored Group Field Theory graphs in arbitrary dimension.
\end{abstract}

\section{Introduction}

Discrete structures over finite sets, in particular graphs, are paramount to our present understanding of physics. Since Feynman realized that the perturbation series of quantum field theory is indexed by subclasses of graphs, the best experimentally tested physical predictions we have to this date rely solely on them. 

Different quantum field theories generate different classes of graphs. The scalar $\Phi^4$ field theory generates graphs formed of four valent vertices and lines. More involved quantum field theories, like Yang-Mills gauge theories \cite{Naka,IZ}, require further structure to be added (new particles, space-time indices, etc.). Random matrix models \cite{mm,gross,Sasakura:1990fs} and non commutative quantum field theories \cite{Conn1,DouNe} generate ribbon graphs. A striking feature of the random matrix models and non commutative quantum field theories \cite{GrWu1,GrWu2,RVW,GMRV,noi} is that the graphs are organized hierarchically. That is, the dominant contribution to the partition function is given by planar graphs, first order corrections are given by genus one graphs, second order corrections by genus two graphs, etc.

Random matrix models are relevant to very diverse physical and mathematical questions ranging from two dimensional quantum gravity to knot theory and quark confinement \cite{Hoo} . In the context of non commutative quantum field theory the topological power counting of the ribbon graphs has been shown in a series of papers to lead to a non trivial fixed point of the renormalization group flow \cite{GrWubeta,DR1,DGMR,GuRo,GGR}. One can therefore expect that an appropriate generalization of such models to higher dimensions should also pose non trivial renormalization fixed points. The study of such generalizations holds essential clues for problems ranging from the quantization of gravity in higher dimensions to condensed matter.

Random matrix models generalize in higher dimensions to random tensor models, or group field theories (GFT) \cite{boulatov,laurentgft,iogft}. The perturbative development of such theories generates ``stranded graphs \cite{DP-P}.''. The connection between GFTs and quantum gravity has been largely investigated \cite{barrett}. Different models have been considered \cite{EPR,Etera,FK}, and their semiclassical limit analyzed \cite{semicl,gravit}. The study of the renormalization properties of such models has been started \cite{FGO,Magnen:2009at, Malek}. However, classical GFT models generate many singular graphs (that is graphs whose dual topological spaces have extended singularities). In a previous paper \cite{Color} we proposed a solution to this problem in the form of the ``colored group field theory'' (CGFT). The singular graphs are absent in this context in any dimension and the surviving graphs possess a cellular complex structure. 

In this paper we extend the study started in \cite{Color} of the Feynman graphs of the CGFT to open graphs (that is graphs with external half lines). For every such graph $\cG$ we define its boundary graph $\cG_{\partial}$. We prove that $\cG_{\partial}$ has a cellular structure inherited from the graph $\cG$. Extending the definition of the boundary operator of \cite{Color}, we introduce the homology of $\cG_{\partial}$ and explore some of its properties. Our model has been further studied in
\cite{Geloun}.

A simple and yet powerful way to encode information about a graph is through topological polynomials. Introduced first by Kirchhoff \cite{kirchhoff} they were studied (much) later by Tutte \cite{Tutte} as the solution of an inductive contraction deletion equation. The topological polynomials appear naturally in the dimensional regularization of quantum field theories \cite{HV} or in the study of statistical physics models \cite{Crapo,Sokal1,Sokal2}. The Tutte polynomials have been generalized by Bollob\'as and Riordan \cite{BR1,BR2,EM1,EM2} to ribbon graphs. Further generalizations of these polynomials, respecting more involved induction equations, have been put in relation with the Feynman amplitudes of random matrix models and non commutative quantum field theories \cite{GurauRiv,RivTan,param-GMRT,KRTW}. 

Relying on the cellular complex structure of $\cG$ and $\cG_{\partial}$ we propose a generalization of the classical topological polynomials adapted to CGFT graphs. These polynomials respect a contraction deletion equation and encode information about the cellular homology of the CGFT graph.

This paper is organized as follows. In section \ref{sec:tut} we briefly review the classical Tutte and Bollob\'as-Riordan polynomials. In section \ref{sec:graphs} we detail the GFT graphs and define the boundary cellular complex and cellular homology for open graphs. In section \ref{sec:poly} we define the topological polynomials of CGFT graphs and show that they obey a contraction deletion relation. Section \ref{sec:conclu} draws the conclusions of our work.

The mathematics and physics nomenclature for graphs is very different and sometimes quite confusing. The reader is strongly encouraged to consult \cite{KRTW} for a dictionary. Also, some familiarity with ribbon graphs is assumed. Again \cite{KRTW} (specifically sections 4.1 and 4.3) provides a very good and concise introduction to this topic.

\section{Tutte and Bollob\'as-Riordan polynomials}
\label{sec:tut}

This section is a short introduction to topological graph polynomials, see \cite{KRTW} and references therein for more detailed presentations. 

A graph $\cG$ is defined by the sets of its vertices $\cV(\cG)$ and lines $\cL(\cG)$. A line, connecting the vertices $v_1, v_2\in \cV(\cG)$ is denoted $l_{v_1v_2}\in \cL(\cG)$.
For any line $l_{v_1v_2}$ of $\cG$ one can define two additional graphs\footnote{The two end vertices might coincide, $v_1=v_2$.}
\begin{itemize}
\item The graph with the line $l_{v_1v_2}$ {\it deleted}, denoted $\cG-l_{v_1v_2}$, with set of lines  $\cL(\cG-l_{v_1v_2})=\cL(\cG)\setminus \{ l_{v_1v_2} \}$ and set of vertices  $\cV(\cG-l_{v_1v_2})=\cV(\cG)$.
\item The graph with the line $l_{v_1v_2}$ {\it contracted}, denoted $\cG/l_{v_1v_2}$, is the graph obtained from $\cG$ by deleting $l_{v_1v_2}$ and identifying the two end vertices $v_1$ and $v_2$. That is $\cL(\cG/l_{v_1v_2})=[\cL(\cG)\setminus \{l_{v_1v_2} \}]/(v_1\sim v_2)$, $\cV(\cG/l_{v_1v_2})=\cV(\cG)/(v_1\sim v_2)$.
\end{itemize}

Note that if $v_1=v_2$ the $\cG/l_{v_1v_2}=\cG-l_{v_1v_2}$. 

Given a graph $\cG$ one can consider the family of its subgraphs. $\cH$ is a subgraph of 
$\cG$ (denoted $\cH\subset \cG$) if $\cV(\cH)=\cV(\cG)$ and $\cL(\cH)\subset \cL(\cG)$. Thus $\cG-l_{v_1v_2}$ is a subgraph of $\cG$, whereas $\cG/l_{v_1v_2}$ is not.

The multivariate Tutte polynomial $Z_{\cG}(q,\{\beta\})$ of the graph $\cG$ depends on one variable $\beta_{l_{v_1v_2}}$ associated to each line $l_{v_1v_2}$ and an unique variable $q$ counting the connected components of $\cG$

\begin{definition}[Sum over subgraphs]\label{def:Tutte}
\begin{equation}\label{eq:defTutte}
Z_{\cG}(q,\{\beta\})=\sum_{\cH\subset  \cG}q^{|k(\cH)|}\prod_{l_{v_1v_2}\in \cL(\cH)}\beta_{l_{v_1v_2}} ,
\end{equation}
where $k(\cH)$ is the number of connected components of the subgraph $\cH$.
\end{definition}
This polynomial obeys a contraction deletion equation
\begin{lemma}\label{lem:condel}
For any line $l_{v_1v_2}\in \cL(\cG)$,
\be\label{eq:contrTutte}
Z_{\cG}( q,\{ \beta \} )= \beta_{l_{v_1v_2}} Z_{\cG/l_{v_1v_2}} (q, \{\beta\}\setminus \{\beta_{l_{v_1v_2}}\} ) + 
Z_{\cG-l_{v_1v_2}} (q, \{ \beta\}\setminus \{ \beta_{l_{v_1v_2}} \} ) .
\ee
For a graphs with no lines but with $v$ vertices $Z_{\cG}(q,\emptyset)= q^v$.
\end{lemma}

In quantum field theory one deals with graphs whose vertices are furthermore decorated with ``half lines'', or external legs\footnote{Or flags in the mathematical literature.}. We use halflines to encode information about the graph $\cG$ in a subgraph $\cH$. We will {\it always} replace a line belonging to $\cG$ but not to $\cH$ by two halflines on its end vertices.

\begin{figure}[htb]
\begin{center}
 \includegraphics[width=3cm]{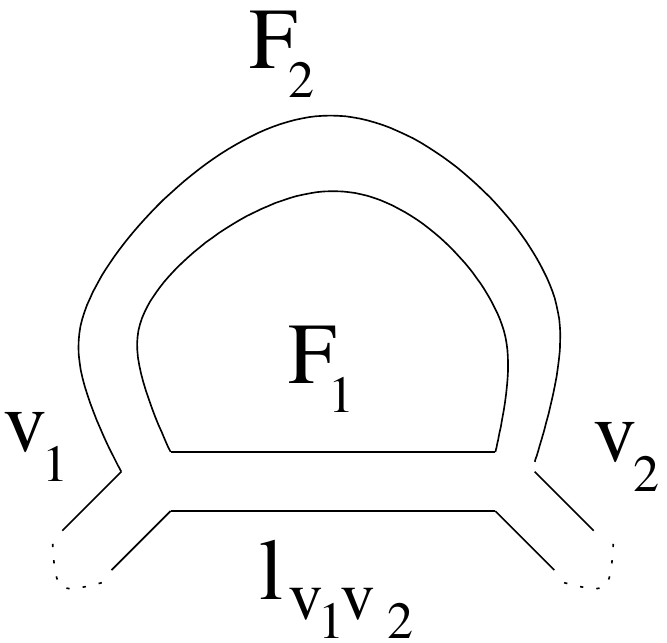}
\caption{Ribbon vertices, ribbon lines and strands}
\label{fig:ribbon}
\end{center}
\end{figure}

The Tutte polynomial can be generalized to ribbon graphs.
A typical ribbon graph with half lines is presented in figure \ref{fig:ribbon}. It is made of ribbon vertices ($v_1$ and $v_2$ in figure \ref{fig:ribbon}) and ribbon lines ($l_{v_1v_2}$ in figure \ref{fig:ribbon}). The lines and half lines in a ribbon graph have two sides, also called {\it strands}, represented by solid lines in figure  \ref{fig:ribbon}.

The strands of a graph encode an extra structure. Tracing a strand one encounters one of the two cases
\begin{itemize}
 \item Either one does {\it not} encounter a half line ($F_1$ in figure \ref{fig:ribbon}).
In this case the closed strand defines an {\it internal face}.
\item Or one {\it does} encounter a half line ($F_2$ in figure \ref{fig:ribbon}). In this case one continues on the second strand of this external half line (one ``pinches'' the external half line). The strands thus traced define an {\it external face}.
\end{itemize}

This ``pinching'' is represented by the dotted curves in figure \ref{fig:ribbon}. 

A ribbon subgraph $\cH\subset \cG$ of the ribbon graph $\cG$ has the same set of vertices 
$\cV(\cH)=\cV(\cG)$, but only a subset of the lines $\cL(\cH)\subset \cL(\cG)$. Again, for a subgraph $\cH$ all lines $l_{v_1v_2}\in \cL(\cG)\setminus \cL(\cH)$ are replaced by {\it pinched} external half lines. Thus, all internal faces of $\cH$ are internal faces of $\cG$, but there might exist external faces of $\cH$ consisting of the union of pieces belonging to several internal faces of $\cG$.

We are now in position to generalize the definition \ref{def:Tutte} to ribbon graphs. We introduce an extra variable $z$ counting {\it all} the faces (internal or external) of the graph, and define
\begin{definition}\label{def:BR}
The multivariate Bollob\'as-Riordan polynomial of a
ribbon graph, analog to the multivariate polynomial of eq. (\ref{eq:defTutte} ), is:
\begin{equation}\label{eq:BR}
V_{\cG}(q,\{\beta_l\},z)=\sum_{\cH\subset \cG} q^{k(\cH)}
(\prod_{l_{v_1v_2}\in \cL(\cH)}  \beta_{l_{v_1v_2}}) \, z^{F(\cH)},
\end{equation}
where $k(\cH)$ is again the number of connected components of $\cH$, and $F(\cH)$ the total number of faces.
\end{definition}

The deletion of a ribbon line $l_{v_1v_2}$ consists in replacing it by two {\it pinched} halflines on its end vertices $v_1$ and $v_2$. It is well defined for all the lines of a graph. On the contrary, the contraction must respect the strand structure and is well defined only for lines $l_{v_1v_2}$ connecting two {\it different} vertices $v_1\neq v_2$. The polynomial define by equation (\ref{eq:BR}) respects the contraction deletion equation (\ref{eq:contrTutte}) only for such lines. The  end graphs (those which can not be contracted further) consist of connected components with only one vertex, but possibly many lines and faces. The polynomial of such end graphs can be read from equation (\ref{eq:BR}).

The crucial property of the topological polynomials is that the definitions in term of subgraphs and the contraction deletion properties can be exchanged. That is, the polynomials of definitions \ref{def:Tutte} and \ref{def:BR} are the {\it unique} solutions of the deletion contraction equation
(\ref{eq:contrTutte}) respecting the appropriate forms for the end graphs. Although, given just the equation (\ref{eq:contrTutte}), one might think that its solution depends on the order in which the lines are contracted (deleted), the equations (\ref{eq:defTutte}) and (\ref{eq:BR}) show that it does not.

\section{Colored Group Field Theory Graphs}
\label{sec:graphs}

Ribbon graphs generalize in higher dimensions to group field theory graphs \cite{laurentgft,iogft,DP-P}. The GFT graphs are generated by a path integral and are built by the following rules. 

The GFT vertex in $n$ dimension has coordination $n+1$. Each halfline (and consequently line) has exactly $n$ strands. Inside a vertex, the strands connect two half lines. In $n$ dimensions, if we label the strands of a halfline 1 to $n$ turning {\it anticlockwise}, the strand $p$ connects to the $p$'th successor halfline when turning {\it clockwise} around the vertex.
Every GFT line connects two half lines with an arbitrary permutation of the strands.

\begin{figure}[htb]
\begin{center}
 \includegraphics[width=4cm]{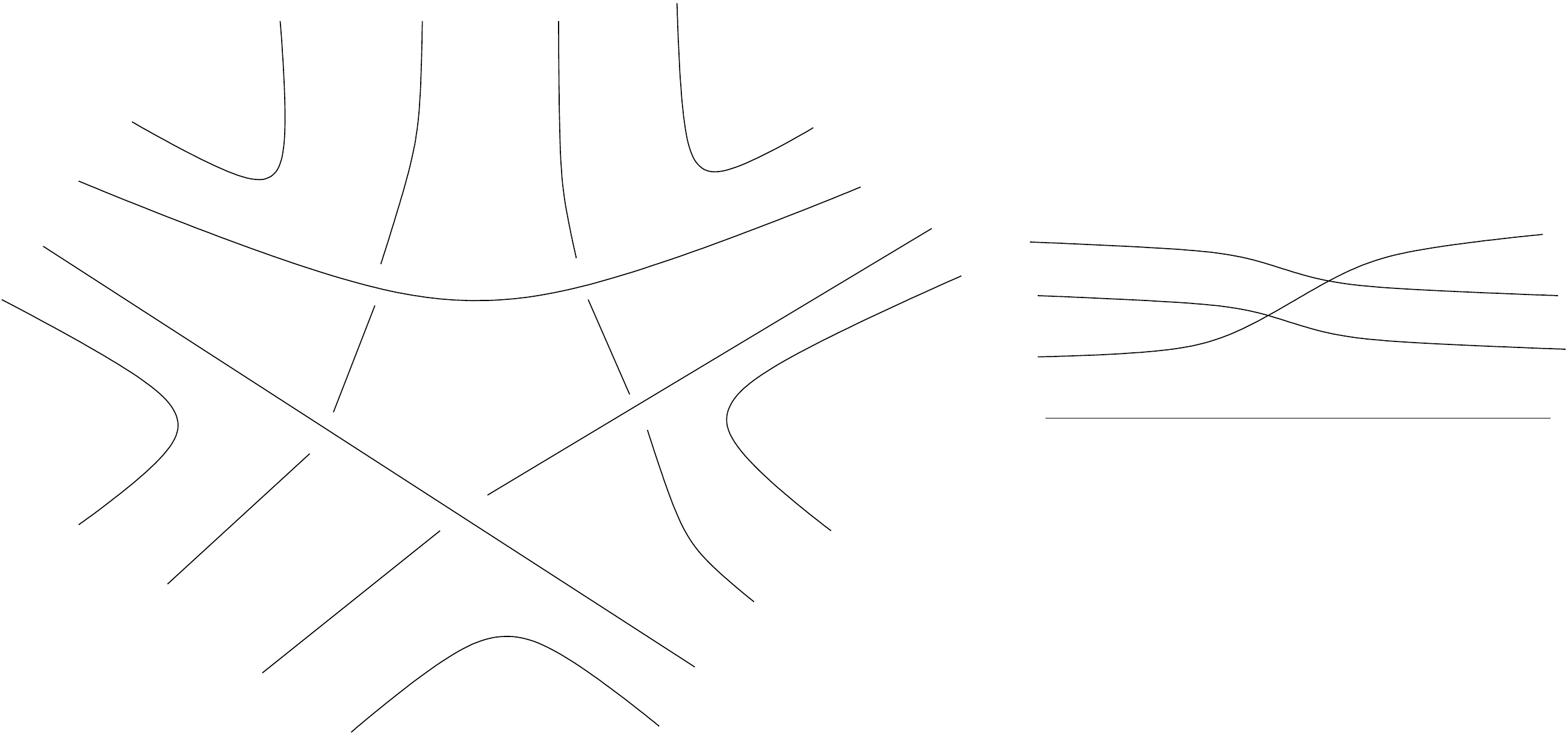}
\caption{GFT vertex and a GFT line in four dimensions.}
\label{fig:4dim}
\end{center}
\end{figure}

Figure \ref{fig:4dim} presents the GFT vertex and a typical GFT line in 4 dimensions.

The reader can check that a GFT graph in two dimensions is a ribbon graph with vertices of coordination three. As such, it is dual to a triangulation of a two dimensional surface. Considering the ribbon vertices of the graph as 0-cells, its lines as 1-cells and its faces as 2-cells, a ribbon graph becomes a two dimensional cellular complex. One would expect that the GFT graphs in higher dimension also have a cellular complex structure. This is not true in general because the permutations of strands on the lines prevent one from defining cells of dimension higher that two!

A solution is to consider only the colored group field theory graphs introduced in \cite{Color}. In fact, to our knowledge, this is the {\it only} category of graphs generated by a path integral which has an associated complex structure in arbitrary dimension\footnote{In three dimensions one also has the alternative to use the orientable model of \cite{FGO}, but this can not be generalized to higher dimensions.}! The graphs obtained by the perturbative development of the color group field theory action of \cite{Color} obey

\begin{definition}\label{def:cgr}
 A CGFT graph in $n$ dimensions is a GFT graph such that
\begin{itemize}
 \item The CGFT vertices are stranded vertices. The set of vertices $\cV(\cG)=\{v_1,\dots v_n\}$ is 
the disjoint union of two sets $\cV(\cG)=\cV^+(\cG)\cup \cV^-(\cG)$. $\cV^+(\cG)$ is the set of positive vertices and  $\cV^-(\cG)$ is the set of negative vertices.
\item The lines $l^i_{v_1v_2}\in \cL(\cG)$ connect a positive and a negative vertex ($v_1\in \cV^+(\cG)$ and $v_2\in \cV^-(\cG)$) and posses a color index $i\in \{0,\dots n \}$. The $n$ strands of all CGFT lines are parallel. Halflines also possess a color index.
\item Each color appears exactly once among the lines or halflines touching a vertex. The colors are encountered in the order $0,\dots, n$ when turning clockwise around a positive vertex and anticlockwise around a negative one. 
\end{itemize}
\end{definition}

A CGFT graph admits two equivalent representations, either as a stranded graph, or simply as an edge  colored graph, obtained by collapsing all the strands belonging to all lines. As the connectivity of strands inside the CGFT vertex and lines are fixed the two representations are in one to one correspondence. 

\begin{figure}[htb]
\begin{center}
 \includegraphics[width=3cm]{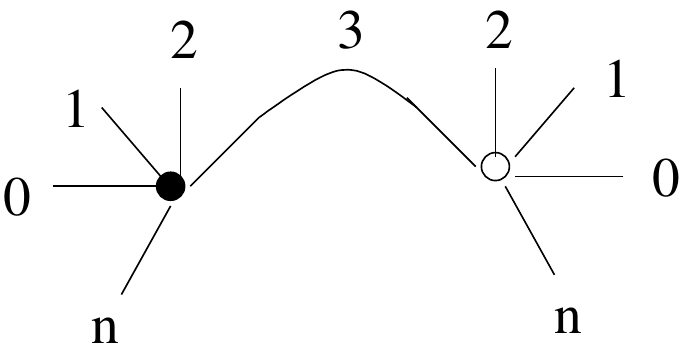}
\caption{Two vertices and a line in a colored graph.}
\label{fig:color}
\end{center}
\end{figure}

A colored graph is made of colored lines connecting positive and negative vertices. In figure 
\ref{fig:color}, the line of color $3$ connects the positive vertex on the left with the negative one on the right. Figure \ref{fig:colorex} gives the two representations for the same graph.

\subsection{Bubbles and cellular structure}

In the definition of the  Bollob\'as-Riordan polynomial the faces (internal and external) played a crucial role. In higher dimensions the faces generalize to higher dimensional cells, called bubbles.

First consider $\cG$ a CGFT graph with no external half lines. In \cite{Color} we defined the $p$-cells of $\cG$ as
\begin{definition}\label{def:bulkbub}
A ``p-bubble'' with colors $i_1<\dots <i_p$ of a graph with $n+1$ colors $\cG$ with no external halflines is a maximal connected components made of lines of colors $i_1,\dots, i_p$. We denote it $\cB^{\cC}_{\cV}$, where $\cC=\{i_1,\dots,i_p\}$ is the ordered set of colors of the lines in the bubbles and $\cV$ is the set of vertices.
\end{definition}

\begin{figure}[htb]
\begin{center}
 \includegraphics[width=7cm]{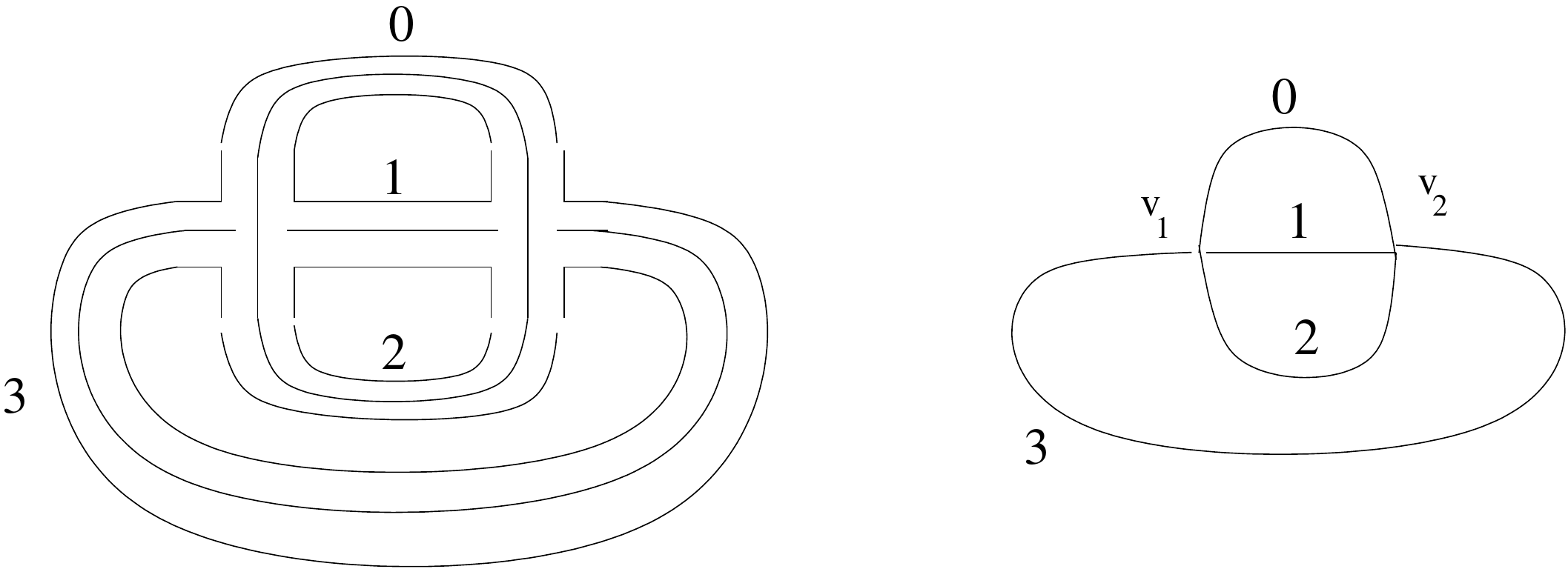}
\caption{A closed colored graph in 3 dimensions.}
\label{fig:colorex}
\end{center}
\end{figure}

Note that, unlike the subgraphs of section \ref{sec:tut}, the connected components {\it do not} have half lines. For example, for the graph in figure \ref{fig:colorex} we have the  3-bubbles $\cB^{012}_{v_1v_2}$, $\cB^{013}_{v_1v_2}$, $\cB^{023}_{v_1v_2}$ and $\cB^{123}_{v_1v_2}$, the 2-bubbles (that is faces) $\cB^{01}_{v_1v_2}$, $\cB^{02}_{v_1v_2}$, $\cB^{03}_{v_1v_2}$, $\cB^{12}_{v_1v_2}$, $\cB^{13}_{v_1v_2}$, $\cB^{23}_{v_1v_2}$, the one bubbles (that is lines) $\cB^0_{v_1v_2}$, $\cB^1_{v_1v_2}$, $\cB^2_{v_1v_2}$, $\cB^3_{v_1v_2}$, and finally the 0-bubbles (that is vertices) $\cB_{v_1}$, $\cB_{v_2}$.

Like the graph $\cG$, the $p$-bubbles themselves admit graphical representations either as stranded graphs or as edge colored graphs. For instance in figure \ref{fig:colorex}, the stranded graph of the $3$-bubble $\cB^{012}_{v_1v_2}$ is obtained by deleting all strands belonging to the line $l^3_{v_1v_2}$. Similarly the stranded graph of the $2$-bubble $\cB^{01}_{v_1v_2}$ is obtained by deleting all strands belonging to the lines $l^{2}_{v_1v_2}$ and $l^3_{v_1v_2}$.

Considering the representation of bubbles as stranded graphs it is easy to see that in any dimension, the {\it strands} themselves always correspond to $2$-bubbles. This remark is crucial for the next section.

As proved in \cite{Color}, the $p$-bubbles define a cellular complex and a cellular homology induced by the boundary operator

\begin{definition}\label{def:bound}
The $p$'th boundary operator $d_p$ acting on a $p$-bubble $\cB^{\cC}_{\cV}$ 
 with colors $\cC=\{i_{1},\dots i_{p}\}$ is
\begin{itemize} 
\item for $p\ge 2$,
\bea \label{eq:boundary}
d_p(\cB^{\cC}_{\cV})=
\sum_{q}(-)^{q+1} \sum_{ \stackrel{ {\cal B'}^{\cC'}_{\cV'}\in \mathfrak{B}^{p-1} } 
{ \cV'\subset \cV \; \cC' = \cC \setminus i_{q}} }  
{\cal B'}^{\cC'}_{\cV'} \;,
\eea
which associates to a $p$-bubble the alternating sum of all $(p-1)$-bubbles formed by subsets of its vertices. 

\item for $p=1$, as the lines $\cB^i_{v_1v_2}$ connect a positive vertex ($v_1\in \cV^+(\cG)$) to a negative one, $v_2\in \cV^-(\cG)$
\bea
 d_1 \cB^i_{v_1v_2}= \cB_{v_1}-\cB_{v_2} \;.
\eea
 \item for $p=0$, $d_0\cB_v=0$.
\end{itemize}
\end{definition}

\subsection{External Half Lines and the Boundary Complex}

A graph $\cG$ with external half lines is dual to a topological space with boundary. We will first associate to $\cG$ a ``boundary graph'' $\cG_{\partial}$, dual to a triangulation of the boundary of the topological space and then identify a cellular complex structure for $\cG_{\partial}$.

To understand the construction of $\cG_{\partial}$ one needs to consider the topological space dual to $\cG$ (see \cite{Color} and \cite{FGO} for details). The dual of a colored graph is essentially a simplicial complex\footnote{It is in fact a slightly more general gluing of simplices along there faces.}. Each CGFT vertex is dual to a $n$-simplex $\Delta^{n}$. The half lines of a vertex are dual to the ``sides'' of $\Delta^n$, that is the $(n-1)$-simplices $\Delta^{n-1}$ bounding it. A boundary simplex $\Delta^{n-1}$ inherits the color of the halfline to which it coresponds. The lines (which are identifications of halflines) corespond to the gluing of the two $\Delta^n$ simplices along a common $\Delta^{n-1}$ boundary simplex.
Higher dimensional $p$-bubbles are dual to $(n-p)$-simplices, in particular the $2$-bubbles are dual to $\Delta^{n-2}$ simplices. In particular, in the stranded representation of a CGFT graph, the $\Delta^{n-2}$ simplices are dual to the strands.

\begin{figure}[htb]
\centering
\includegraphics[width=3cm]{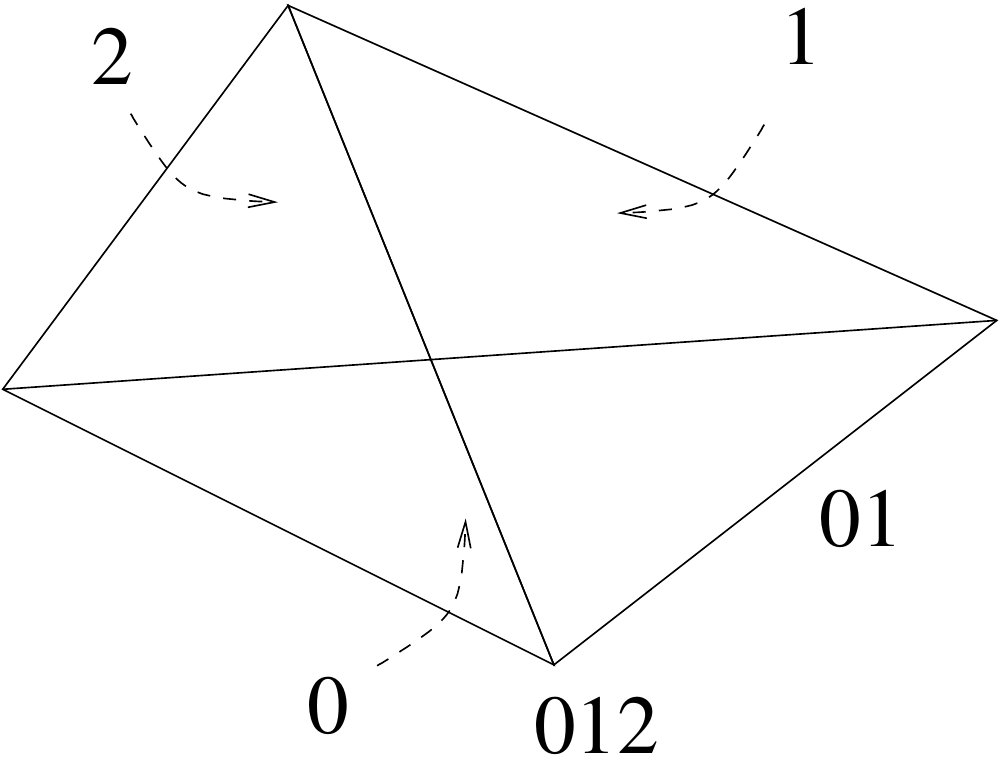}
\caption{Tetrahedron dual to a CGFT vertex.}
\label{fig:tetrad}
\end{figure}

In three dimensions this is represented in figure \ref{fig:tetrad}. The vertex $0123$ is dual to the tetrahedron $0123$, the halfline $0$ is dual to the triangle $0$, the 2-bubble $01$ is dual to the edge common to the triangles $0$ and $1$, and the 3-bubble $012$ is dual to the the vertex of the tetrahedron common to the triangles $0$, $1$, and $2$. 

If a vertex in a CGFT graph has no half lines then its dual simplex $\Delta^n$ sits in the interior of the simplicial complex (in the bulk). On the contrary, if a vertex has half lines, then its dual simplex sits on the boundary of the simplicial complex, and contributes to the triangulation of this boundary with the $\Delta^{n-1}$ simplex dual to the half line. The triangulation of the boundary of the simplicial complex is therefore made of all the $\Delta^{n-1}$ simplices dual to the halflines of the graph. These $\Delta^{n-1}$ simplices are glued along there boundary $\Delta^{n-2}$. The boundary $\Delta^{n-2}$ simplices are dual, in the stranded representation of a CGFT graph to the open strands.

\begin{figure}[htb]
\centering
\includegraphics[width=6cm]{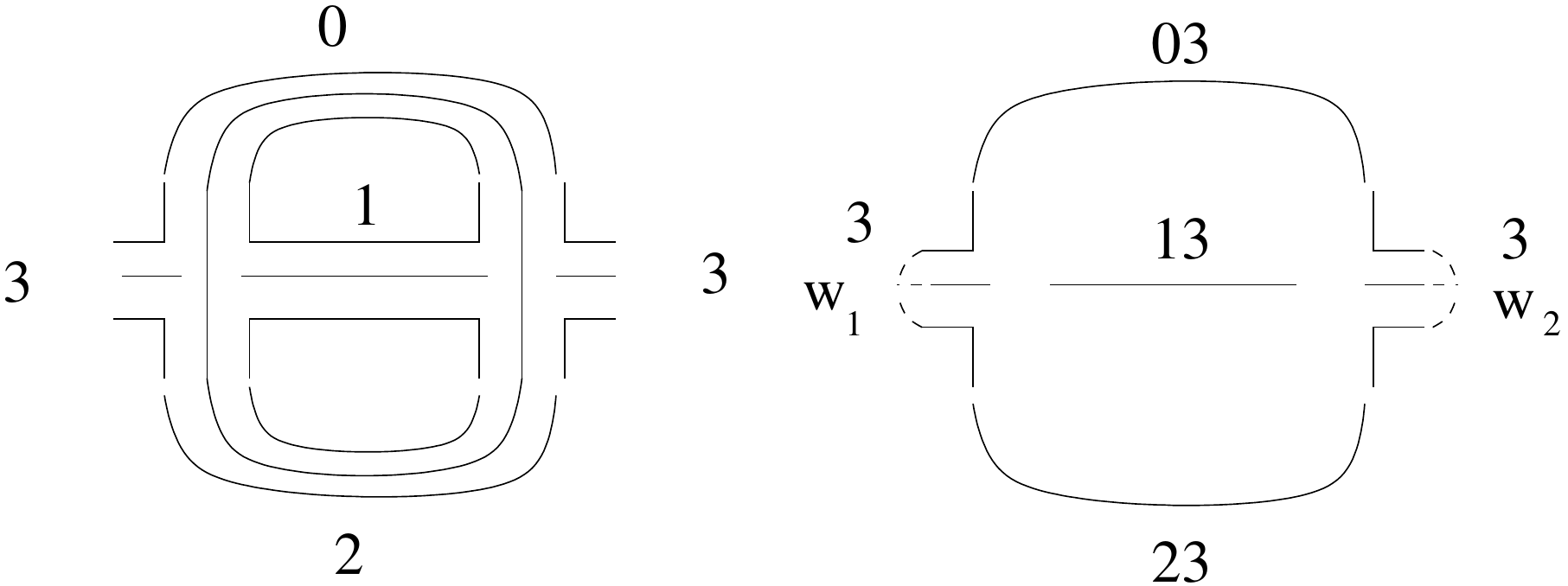}
\caption{A CGFT graph $\cG$ and it boundary graph $\cG_{\partial}$.}
\label{fig:pinch}
\end{figure}

To obtain the graph $\cG_{\partial}$ dual to the boundary of the simplicial complex one must draw a vertex for each external halfline of $\cG$ and a line for each open strand of $\cG$. This can be achieved starting with the stranded representation of the graph $\cG$ (see figure \ref{fig:pinch}), delete all closed strands, and ``pinch'' the external strands into a vertex for each external half line. The graph thus obtained is the edge colored representation of $\cG_{\partial}$. We call $\cG_{\partial}$ the ``boundary graph'' of $\cG$.

The vertices of $\cG_{\partial}$ inherit the color of the halfline and the lines of $\cG_{\partial}$ inherit the {\it couple} of colors of the strand to which they corespond.  In the example of figure \ref{fig:pinch}, the graph $\cG_{\partial}$ (represented on the right) has one connected component with two vertices, $w_1$ and $w_2$, both of color $3$ and three lines of colors $03$, $13$ and $23$.

Note that $\cG_{\partial}$ is a graph of vertices with one color and lines colored by couples of colors: a priori it is very different from a CGFT graph. Nevertheless $\cG_{\partial}$ has a cellular complex structure, strongly reminiscent of the one of $\cG$. We denote the set of vertices of $\cG_{\partial}$ (obtained after pinching) by $\cV_{\partial}$. They are the $0$-bubbles of the cellular complex of $\cG_{\partial}$. For $p\ge 1$, we have
\begin{definition}\label{def:boundbub}
 Let a graph $\cG$ and its boundary graph $\cG_{\partial}$ obtained after pinching. For $p\ge 1$ the
``boundary  $p$-bubbles'' $(\cB_{\partial})^{\cC'}_{\cV'_{\partial}}$ are the maximally connected components of $\cG_{\partial}$ formed by boundary vertices $\cV'_{\partial}\subset \cV_{\partial}$ and boundary lines of colors $i_ai_b$, with $\{i_a, i_b\}\subset \cC'\subset \{0,\dots n\}$ and $|\cC'|=p+1$.
\end{definition}

For example $\cG_{\partial}$ in figure \ref{fig:pinch} has 
\begin{itemize}
\item $0$ bubbles $(\cB_{\partial})^{3}_{w_1}$, $(\cB_{\partial})^{3}_{w_2}$, which are the vertices of $\cG_{\partial}$.
\item $1$ bubbles $(\cB_{\partial})^{03}_{w_1w_2}$, $(\cB_{\partial})^{13}_{w_1w_2}$, $(\cB_{\partial})^{23}_{w_1w_2}$, which are the lines of $\cG_{\partial}$.
\item $2$ bubbles $(\cB_{\partial})^{013}_{w_1w_2}$, $(\cB_{\partial})^{023}_{w_1w_2}$, $(\cB_{\partial})^{123}_{w_1w_2}$, which are the connected components with lines $(03,13)$, $(03,23)$  and $(13,23)$ respectively.
\end{itemize}
We denote $\mathfrak{B}^p_{\partial}$ the set of all boundary $p$-bubbles, and following \cite{Color} we define the operator
\begin{definition}\label{def:boundbound}
The $p$'th boundary operator $d^{\partial}_p$ of the boundary complex, acting on a boundary $p$-bubble $(\cB_{\partial})^{\cC}_{\cV_{\partial}}$ 
 with colors $\cC=\{i_{1},\dots i_{p+1}\}$ is
\begin{itemize} 
\item for $p\ge 1$,
\bea \label{eq:boundboundary}
d^{\partial}_p[(\cB_{\partial})^{\cC}_{\cV_{\partial}}]=
\sum_{q}(-)^{q+1} \sum_{ \stackrel{ {(\cB'_{\partial})}^{\cC'}_{\cV'_{\partial}}\in \mathfrak{B}^{p-1}_{\partial} } 
{ \cV'_{\partial}\subset \cV_{\partial} \; \cC' = \cC \setminus i_{q}} }  
(\cal B'_{\partial})^{\cC'}_{\cV'_{\partial}} \;,
\eea
 \item for $p=0$, $d^{\partial}_0[(\cB_{\partial})^{i_1}_w]=0$.
\end{itemize}
\end{definition}

For $\cG_{\partial}$ of figure \ref{fig:pinch} for instance,
\bea
  d_2^{\partial}[(\cB_{\partial})^{013}_{w_1w_2}]&=&(\cB_{\partial})^{13}_{w_1w_2}
-(\cB_{\partial})^{03}_{w_1w_2}\nonumber\\
 d_1^{\partial}[d_2^{\partial}[(\cB_{\partial})^{013}_{w_1w_2}]]&=&(\cB_{\partial})^{3}_{w_1} + 
(\cB_{\partial})^{3}_{w_2} - (\cB_{\partial})^{3}_{w_1}
-(\cB_{\partial})^{3}_{w_2}=0 \; .
\eea
The operator $d_p^{\partial}$ is a boundary operator in the sense
\begin{lemma} 
 \bea
  d^{\partial}_{p-1} \circ d^{\partial}_p =0 \; .
 \eea
\end{lemma}
{\bf Proof:}
The proof goes much like its counterpart presented in \cite{Color}. Consider the application of two consecutive boundary operators on a boundary $p$-bubble
\bea
&&d^{\partial}_{p-1}d^{\partial}_p[(\cB_{\partial})^{\cC}_{\cV_{\partial}}] = \sum_{q}(-)^{q+1} \sum_{ \stackrel{ {(\cal B'_{\partial})}^{\cC'}_{\cV'_{\partial}}\in \mathfrak{B}^{p-1}_{\partial} } 
{ \cV'_{\partial}\subset \cV_{\partial} \; \cC' = \cC \setminus i_{q}} }  d^{\partial}_{p-1}[(\cB'_{\partial})^{\cC'}_{\cV'_{\partial}}]
\\
&&=\sum_{q}(-)^{q+1}
\sum_{ \stackrel{ {(\cal B'_{\partial})}^{\cC'}_{\cV'_{\partial}}\in \mathfrak{B}^{p-1}_{\partial} } 
{ \cV'_{\partial}\subset \cV_{\partial} \; \cC' = \cC \setminus i_{q}} } 
\Big{[}
\sum_{r<q} (-)^{r+1}\sum_{ \stackrel{ {(\cal B''_{\partial})}^{\cC''}_{\cV''_{\partial}}\in \mathfrak{B}^{p-2}_{\partial} } 
{\cV''_{\partial}\subset \cV_{\partial} \; \cC'' = \cC\setminus i_{q} \setminus i_{r}} } 
(\cal B''_{\partial})^{\cC''}_{\cV''_{\partial}}
 +\nonumber\\
&&\sum_{r>q} (-)^{r}
\sum_{ \stackrel{ {(\cal B''_{\partial})}^{\cC''}_{\cV''_{\partial}}\in \mathfrak{B}^{p-2}_{\partial} } 
{\cV''_{\partial}\subset \cV_{\partial} \; \cC'' = \cC\setminus i_{q} \setminus i_{r}} } 
(\cal B''_{\partial})^{\cC''}_{\cV''_{\partial}} \; \;,
\eea
as $i_{r}$ is the $r-1$'th color of $\cC'\setminus i_q$ if $q<r$. The two terms cancel by exchanging  $q$ and $r$ in the second term.

\qed

The boundary bubbles define a cellular complex with attaching maps induced by the boundary operator of definition \ref{def:boundbound}. With the appropriate substitutions, one reproduces the main results of \cite{Color} for the cellular homology of $\cG_{\partial}$ defined by $d_p^{\partial}$.
\begin{lemma} Let $\cG_{\partial}$ a connected boundary CGFT graph with $n+1$ colors. The operator $d_p^{\partial}$ has the following properties
\begin{itemize}
 \item  The $d_0^{\partial}$ operator respects
\bea
\mathrm{ker} (d_0^{\partial}) = \bigoplus_{|\mathfrak{B}^0_{\partial}|}\ZZ \; .
\eea
\item The $d_1^{\partial}$ operator respects
\bea
\mathrm{ker} (d_1^{\partial}) 
=\bigoplus_{|\mathfrak{B}^1_{\partial}|-|\mathfrak{B}^0_{\partial}|+1}\ZZ \; ,\qquad \mathrm{Im}(d_1^{\partial}) \bigoplus_{|\mathfrak{B}^0_{\partial}|-1}\ZZ  \; .
\eea
\item The $d_{n-1}^{\partial}$ operator respects
\bea
 \mathrm{ker} (d_{n-1}^{\partial})= \ZZ\; , \qquad \mathrm{Im} (d_{n-1}^{\partial}) \bigoplus_{|\cB^{n-1}_{\partial}|-1}\ZZ \; .
\eea
\end{itemize}
\end{lemma}

In consequence, for all graphs, denoting the homology groups of $\cG_{\partial}$ as $H_q^{\partial}$, we have
\bea
  H_0^{\partial} = \ZZ \; ,\qquad H_{n}^{\partial} = \ZZ \; .
\eea
And if $\cG$ is moreover a three dimensional graph (that is it has four colors), then for each connected component of $\cG_{\partial}$ we have
\bea
 H^{\partial}_0=\ZZ \; , \qquad H^{\partial}_1=\bigoplus_{2g} \ZZ \; ,\qquad H^{\partial}_2=\ZZ \; ,
\eea
that is $\cG_{\partial}$ is a union of tori.

\section{Topological polynomials of GFT graphs}
\label{sec:poly}

Having at our disposal a good definition of bubbles in arbitrary colored graphs we proceed to generalize the topological polynomials to higher dimensional graphs. However one encounters a problem.

There is an incompatibility between the contraction of lines of section \ref{sec:tut} and the colored graphs of definition \ref{def:cgr}. If $\cG$ is a colored graph and $l$ one of its lines, $\cG-l$ is still a colored graph, but $\cG/l$ is not. The vertex obtained by identifying the endvertices of $l$ does not respect the conditions of definition \ref{def:cgr}. But the $p$-bubbles are defined only for colored graphs. It is therefore needed to modify the contraction move to ensure that $\cG/l$ remains a colored graph. This is achieved by slightly enlarging the class of graphs we consider to graphs with active and passive lines.

\begin{definition}
A colored graph with active and passive lines is a colored graph $\cG$ and a partition of the lines $\cL(\cG)$ into two disjoint sets, $\cL(G)=\cL_1(\cG)\cup \cL_2(\cG)$, such that $\cL_2(\cG)$ is a forest\footnote{That is the lines in $\cL_2(\cG)$ do not form loops.}.The lines in the first set, $\cL_1(\cG)$ are called {\it active} lines whereas the lines in the second set $\cL_2(\cG)$ are called {\it passive}.
\end{definition}

Note that a colored graph with no passive lines is just a colored graph in the sense of definition \ref{def:cgr}. For a colored graph with active and passive lines, we define the deletion and contraction {\it only} for the active 
lines $l\in \cL_1(\cG)$ as follows 
\begin{definition}\label{def:actpas}
 For all active lines $l\in\cL_1(\cG)$ we define
\begin{itemize}
\item The graph with the line $l$ deleted, $\cG-l$ with $\cV(\cG-l)=\cV(\cG)$, $\cL_1(\cG-l)=\cL_1(\cG)\setminus \{l\}$ and $\cL_2(\cG-l)=\cL_2(\cG)$.
\item The graph with the line $l$ contracted $\cG/l$ with $\cV(\cG/l)=\cV(\cG)$, $\cL_1(\cG/l)=\cL_1(\cG)\setminus \{l\}$ and $\cL_2(\cG/l)=\cL_2(\cG)\cup \{l\}$.
\end{itemize}
\end{definition}

That is the contraction is reinterpreted as transforming an active lines into a passive one, instead of the identification of the end vertices. Note that one can use the new definitions of $\cG-l$ and $\cG/l$ also for the graphs of section \ref{sec:tut}. Then the equation (\ref{eq:contrTutte}) holds for all active lines and definition \ref{def:Tutte} holds if $\cL_2(\cG)=\emptyset$.

Let $\cG$ be a CGFT graph with $n+1$ colors, and $\cG_{\partial}$ its boundary graph. As before, let $\mB^p, 0\le p\le n$ be the set of all {\it bulk} $p$-cells (defined by \ref{def:bulkbub}), and $\mB_{\partial}^p,0\le p\le n-1$ the set of {\it boundary} $p$-cells (defined by \ref{def:boundbub}). Denote $\mB^{n+1}$ the set of the connected components of $\cG$ and $\mB^n_{\partial}$ the set of connected components of $\cG_{\partial}$.
To define the topological polynomial associated to $\cG$, we introduce a variable $x_p$ counting all the bulk $p$-cells and a variable $y_p$ counting all the boundary $p$-cells. Furthermore, we associate a variable $\beta_l$ to all active lines in $\cG$.
\begin{definition}
 The topological polynomial $P_{\cG}(\{\beta_l\},\{x_p\},\{ y_p \})$ is
\bea\label{eq:polygen}
P_{\cG}(\{\beta_l\}, \{x_p\},\{ y_p \})= \sum_{\cH\subset\cG; \cL_2(\cH)=\cL_2(\cG)} \Big{(}\prod_{l\in\cL_1(\cH)}\beta_l \Big{)}
\prod_{p=0}^{n+1} x_p^{|\mB^p|}\prod_{p=0}^{n} y_p^{|\mB_{\partial}^p|} 
\; .
\eea
\end{definition}

Note that the variables $x_0$ and $x_1$ are redundant: the number of vertices of any subgraph is equal to the number of vertices of the initial graph, thus $x_0^{|\mB^0|}$ is just an overall multiplicative factor and $x_1$ contributes just with a global $x_1^{\cL_2(\cG)}$ multiplicative factor after a uniform rescaling of the line parameters $\beta_l$.
An explicit example is detailed at length in the Appendix.

The polynomial of equation \ref{eq:polygen} has the following behavior under various rescalings 
\bea
&& P_{\cG}(\{\beta_l\}, \{\rho^{(-)^p}x_p\},\{\rho^{(-)^{p+1}} y_p \}) = \rho^{\chi(\cG)}
P_{\cG}(\{\beta_l\}, \{ x_p\},\{ y_p \}) \nonumber\\
&& P_{\cG}(\{\beta_l\}, \{ x_p\},\{\rho^{(-)^{p}} y_p \}) = \rho^{\chi(\cG_{\partial})}
P_{\cG}(\{\beta_l\}, \{ x_p\},\{ y_p \}) \; ,
\eea
with $\chi(\cG)$ and $\chi(\cG_{\partial})$ the Euler characteristics of $\cG$ and $\cG_{\partial}$ respectively. Moreover it respects the contraction deletion relation
\begin{lemma}
 For all active lines $l$
\bea\label{eq:gendelcontr}
P(\{\beta\}, \{x_p\},\{ y_p \})&&=\beta_l P_{\cG/l} (\{\beta\} \setminus \{ \beta_l \}, \{x_p\},\{ y_p \}  ) \nonumber\\
&& +P_{\cG-l} ( \{\beta \}\setminus \{\beta_l\}, \{x_p\},\{ y_p \} ) ,
\eea
\end{lemma}

{\bf Proof: } Note that any active line $l$ divides the subgraphs indexing the sum in \ref{eq:polygen}, $\cH\subset \cG$ with $\cL_2(\cH)=\cL_2(\cG)$, into two families, namely
\be
\cF_{l\in}(\cG)=\{\cH | l\in \cL_1(\cH)\} \quad \cF_{l\notin} (\cG)=\{\cH | l \notin \cL_1(\cH)\}
\; .
\ee
We split (\ref{eq:polygen}) into two terms corresponding to these two families.
All the subgraphs in the first family contain $l$, thus we can factor $\beta_l$ in front of the first term, and reinterpret the line $l$ as a passive line in the graph $\cH/l$. The set of graphs $\cF_{l\in}(\cG)$ is in one to one correspondence to the set of all the subgraphs 
$\cH/l \subset \cG/l$ with $\cL_2(\cH/l)=\cL_2(\cG/l)=\cL_2(\cG)\cup \{l\}$, therefore the first term on the rhs of eq. \ref{eq:gendelcontr} is recovered. The graphs in the second family $\cF_{l\notin}(\cG)$ coincide with the subgraphs of $\cG-l$, and one recovers the second term in 
\ref{eq:gendelcontr}.

\qed

The classical Tutte and Bollob\'as-Riordan polynomials are recovered as limit cases of the higher dimensional polynomial defined here. For the CGFT graphs with three colors (which are trivalent ribbon graphs) equations (\ref{eq:polygen}) and (\ref{eq:BR}) imply
\bea
 P(\{\beta\}, \{1,1,z,q\}, \{1,1,z\})= V_{\cG}(q,\{ \beta \},z) \; ,
\eea
and for an arbitrary CGFT with $\cL_2(\cG)=\emptyset$
\bea
 P(\{\beta\}, \{1,q\},\{ 1 \})=Z_{\cG}(q,\{ \beta \}) \; .
\eea

\section{Conclusion}
\label{sec:conclu}

In this paper we introduced topological polynomials adapted to CGFT graphs, obeying a deletion contraction equation. To each CGFT graph we first associated a boundary graph, and defined and studied its homology. 

The generalized polynomials reproduce the classical ones for certain values of the parameters.
Although the polynomials we define are not the unique generalization one can consider, they already encode nontrivial topological information as seen by the behavior under rescaling of their arguments.
One can for instance consider generalizations, in which instead of associating a unique variable $x_p$ which counts all the $p$-cells, one associates a different variable to each $p$-cell. Such a polynomial would presumably obey a generalized deletion contraction for $p$-cells instead of lines.

\section*{Acknowledgements}

The author would like to thank Vincent Rivasseau for very useful discussions at an early stage of this work.

Research at Perimeter Institute is supported by the Government of Canada through Industry 
Canada and by the Province of Ontario through the Ministy of Research and Innovation.

\section*{Appendix}

In this appendix we detail the topological polynomial and check the contraction deletion relation for the graph in figure \ref{fig:colorex}. The subgraphs of this graph are: the total graph formed by the lines $0123$, subgraphs with three lines $123$,$023$, $013$, $012$, sub graphs with two lines $01$, $02$, $03$, $12$, $13$, $23$, those with one line $0$, $1$, $2$, $3$ and the subgraph with zero lines. The polynomial of the complete graph is then
\bea\label{eq:trea}
 P_{\cG}&& = \beta_{0} \beta_{1} \beta_{2}\beta_{3} \;  x_0^2 x_1^4x_2^6 x_3^4 x_4 \nonumber\\
&&+ ( \beta_{1} \beta_{2}\beta_{3}+\beta_{0} \beta_{2}\beta_{3}+\beta_{0} \beta_{1} \beta_{3}+\beta_{0} \beta_{1} \beta_{2}) \;
x_0^2x_1^3x_2^3x_3x_4y_0^2y_1^3y_2^3y_3
\nonumber\\
&&+(\beta_{0} \beta_{1}+\beta_{0}\beta_{2}+\beta_{0} \beta_{3} 
+ \beta_{1} \beta_{2}+ \beta_{1} \beta_{3}+ \beta_{2}\beta_{3}) \;
x_0^2x_1^2x_2 x_4y_0^4y_1^6y_2^4y_3
\nonumber\\
&&+(\beta_0+\beta_1+\beta_2+\beta_3) \; x_0^2x_1 x_4y_0^6y_1^9y_2^5y_3 \nonumber\\
&&+x_0^2  x_4^2 y_0^8 y_1^{12}y_2^8 y_3^2 \; .
\eea

Consider for instance the contributions of the subgraph $012$, represented in figure \ref{fig:pinch}. It has two vertices, three lines $0$, $1$ and $2$, three internal faces $01$, $02$ and $12$, one internal bubble $012$ and one connected component. This yields a factor $x_0^2x_1^3x_2^3x_3x_4$. Its boundary graph is represented on the right hand side of figure \ref{fig:pinch}. It has two vertices (both colored $3$), three lines colored $03$, $13$ and $23$, three faces, one formed by the lines $01$ and $02$, another one formed by the lines $01$, $03$ and the third one formed by the lines $02$ and $03$, and one connected component, yielding a factor $y_0^2y_1^3y_2^3y_3$. Multiplying the two factors reproduces the coefficient of $\beta_0\beta_1\beta_2$ in equation \ref{eq:trea}

\begin{figure}[htb]
\centering
\includegraphics[width=8cm]{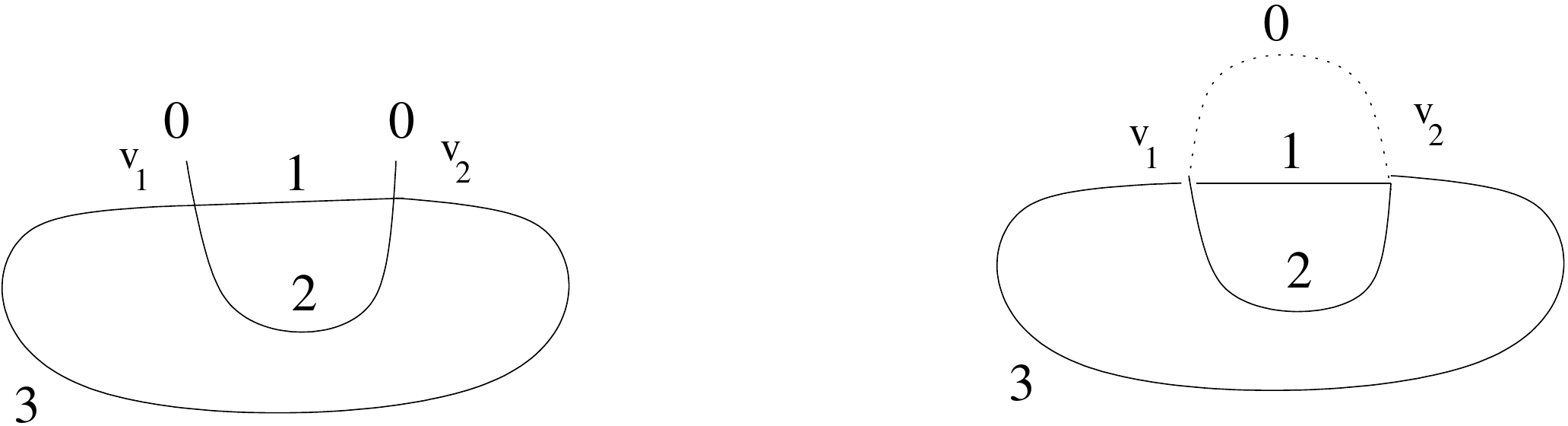}
\caption{The graphs $\cG-l$ and $\cG/l$.}
\label{fig:Gl}
\end{figure}

Chose a line, say $0$. The graphs $\cG-l$ and $\cG/l$ are represented in figure \ref{fig:Gl} where the passive line $l_0$ of $\cG/l$ is represented as a dotted line. 

The graph $\cG-l$ has subgraphs made of lines $123$, $12$, $23$, $13$, $1$, $2$, $3$ and the subgraph with zero lines. Thus
\bea
 P_{\cG-l}=
&& \beta_{1} \beta_{2}\beta_{3} \;
x_0^2x_1^3x_2^3x_3x_4y_0^2y_1^3y_2^3y_3 \nonumber\\
&&+(\beta_{1} \beta_{2}+ \beta_{1} \beta_{3}+ \beta_{2}\beta_{3}) \;
x_0^2x_1^2x_2 x_4y_0^4y_1^6y_2^4y_3
\nonumber\\
&&+(\beta_1+\beta_2+\beta_3) \; x_0^2x_1 x_4y_0^6y_1^9y_2^5y_3 \nonumber\\
&&+x_0^2  x_4^2 y_0^8 y_1^{12}y_2^8 y_3^2 \; .
\eea

All the subgraphs of $\cG/l$ will have $l_0\in \cL_2$ as a passive line. They are formed by the active lines $123$, $12$, $23$, $13$, $1$, $2$, $3$ and the graph with no active line. Therefore

\bea
 P_{\cG/l}&& = \beta_{1} \beta_{2}\beta_{3} \;  x_0^2 x_1^4x_2^6 x_3^4 x_4 \nonumber\\
&&+ ( \beta_{2}\beta_{3}+ \beta_{1} \beta_{3}+ \beta_{1} \beta_{2}) \;
x_0^2x_1^3x_2^3x_3x_4y_0^2y_1^3y_2^3y_3
\nonumber\\
&&+(\beta_{1}+\beta_{2}+\beta_{3} ) \;
x_0^2x_1^2x_2 x_4y_0^4y_1^6y_2^4y_3
\nonumber\\
&&+\; x_0^2x_1 x_4y_0^6y_1^9y_2^5y_3 \; ,
\eea
and direct inspection shows that
\bea
 P_{\cG}= \beta_0 P_{\cG/l}+ P_{\cG-l} \; .
\eea

\end{document}